\documentclass[pra,aps,amsmath,amssymb,reprint,showpacs,floatfix,superscriptaddress,longbibliography]{revtex4-1}
\usepackage{graphicx}
\usepackage{dcolumn}
\usepackage{bm}
\usepackage{mathtools}

\begin{document}


\title{Topological suppression of magnetoconductance oscillations in NS junctions} 



\author{Javier Osca}
\email{javier@ifisc.uib-csic.es}
\affiliation{Institut de F\'{\i}sica Interdisciplin\`aria i de Sistemes Complexos
IFISC (CSIC-UIB), E-07122 Palma de Mallorca, Spain}
\author{Lloren\c{c} Serra}
\affiliation{Institut de F\'{\i}sica Interdisciplin\`aria i de Sistemes Complexos
IFISC (CSIC-UIB), E-07122 Palma de Mallorca, Spain}
\affiliation{Departament de F\'{\i}sica,
Universitat de les Illes Balears, E-07122 Palma de Mallorca, Spain}

\date{January 15, 2017}

\begin{abstract}
We show that the 
magnetoconductance oscillations of laterally-confined 2D NS junctions   
are completely suppressed when the superconductor side enters a topological phase.
This suppression 
can be attributed to the modification of the vortex structure of local currents at the junction 
caused by the  topological transition of the superconductor. The two regimes (with and without oscillations) could be seen in a semiconductor 2D junction with a cleaved-edge geometry, one of the junction arms having proximitized superconductivity.
We predict similar oscillations and suppression as a function of the Rashba coupling.
The 
oscillation suppression is robust against differences in chemical potential and phases of lateral superconductors.
\end{abstract}


\maketitle 

\section{Introduction}

Magnetoconductance oscillations are a central topic of the field of quantum transport
in nanostructures. Famous examples are the celebrated Aharonov-Bohm oscillations 
in a quantum ring and the Shubnikov-deHass oscillations in the quantum Hall effect \cite{ihn}. 
In a general sense, the accumulation of complex phases (angles) in the wave function during orbital motion in a perpendicular magnetic field is the basic cause behind the magnetoconductance oscillations.
In presence of a superconductivity pairing gap, transport can be described as the propagation of electron and hole quasiparticles, with Andreev processes allowing the transformation of one type into the other.
Andreev reflections in a normal-superconductor (NS) junction are affected by a magnetic field
acting on the normal side, the field thus modifying the magnetoconductance of the junction.

The interplay of Andreev reflection and magnetic orbital effects has received a long 
lasting attention \cite{Bee92,Taka98,Asa00,Hop00,Gia05,Ero05,Cht07,Rak07,Kha10,Car13}. 
In particular, 
Takagaki \cite{Taka98} studied the magnetoconductance of a 2D NS junction, laterally confined 
to a width $L_y$ and with the N terminal being a semiconductor. As the magnetic field is increased the magnetoconductance tends
to decrease  in a stepwise manner due to the depopulation of the active Landau bands of the 
semiconductor. It was predicted, however, that for high enough fields conspicuous magnetoconductance
oscillations superimposed to the general stepwise reduction would be present.    
Maxima are related to (electron-hole) Andreev reflection while minima are due to  enhanced normal (electron-electron) reflection. Andreev reflection suppression or enhancement 
appears because of  the spatial separation between the electron and hole edge states, each one attached to a different edge of the N lead. In presence of orbital magnetic effects the only way the two 
edge channels can be connected is through multiple alternating electron-hole and hole-electron reflections
at the transverse boundary of the 2D junction. Therefore the
resulting 
conductance depends on the spatial structure of these reflections with respect to the transverse boundary of the junction. This scenario
of magnetoconductance oscillations 
has been discussed in detail in Refs.\ \cite{Taka98,Taka00,Asa00,Hop00,Hsu04,Gia05,Ero05,Bat07,Cht07,Rak07,Sun08,Kha10,Car13} and it has been experimentally 
confirmed in Ref.\ \cite{Ero05}.

\begin{figure}[t]
\centering\resizebox{0.455\textwidth}{!}{
\includegraphics[width=\columnwidth,clip=true]{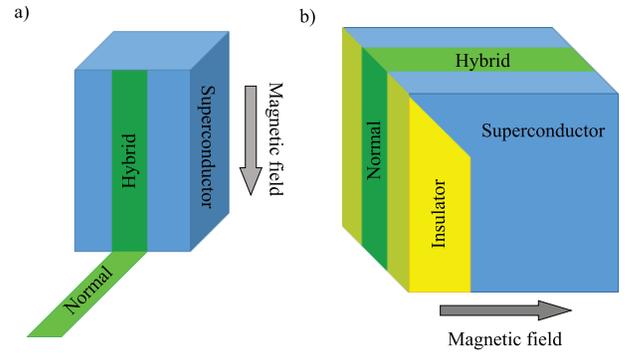}%
}
\caption{Sketches of cleaved-edge 2D wires in a uniform magnetic field. One of the arms is proximity coupled with an $s$-wave superconductor.
The magnetic field is perpendicular and parallel to the normal and hybrid 2D leads, respectively.}
\label{F1}
\end{figure}

In this work we investigate the fate of Takagaki oscillations when the {\em trivial} superconductor
lead
is replaced by a topological superconductor lead. 
We consider a 2D junction of semiconductor wires with effective topological superconductivity 
in one of the junction sides induced by proximity with an $s$-wave superconductor. It has been 
much studied recently how
the resulting hybrid semiconductor-superconductor system can be driven into a topological phase
by the combined action of a parallel magnetic field and Rashba spin-orbit interaction \cite{Fu,Oreg,Lutchyn}. In hybrid nanowires the characteristic of the topological phase 
is the emergence of
zero-energy Majorana modes attached to
possible potential barriers 
or wire ends \cite{Alicea,Leijnse,Beenakker,Franz,StanescuREV}.
In this work, we provide evidence that 
in nanowire junctions
the nature of the Andreev reflections changes from the trivial to the topological phase such that magnetoconductance oscillations are 
completely suppressed in the topological phase.
The abrupt magnetoconductance change of behavior across the phase transition
is thus another clear signal of the topological superconductivity.

The system we have in mind, sketched in Fig.\ \ref{F1}, is a 2D semiconductor wire 
in a cleaved-edge-like geometry \cite{Pfe90}, with one of the two arms proximitized with a 
standard $s$-wave superconductor. The magnetic field is such that it is
perpendicular and parallel to the normal and proximitized arms, respectively. This way we may 
simultaneously achieve with a uniform field
the required quantum Hall behavior on the N arm and the possibility to induce the topological 
transition on the hybrid S (proximitized) arm.
The cleaved-edge device suggested by Fig.\ \ref{F1} assumes a uniform magnetic field, a realistic 
approximation in view of the small size of the nanostructure. In principle, non uniform fields 
could also be created by attaching micromagnets \cite{Kja12},  but this would possibly be a 
technically more involved alternative. It should also be stressed that our simplified model only considers the
different relative orientations of the magnetic field with respect to the quasiparticle motions, while other effects of the cleavage such as confinement inhomogeneities affecting the motion at the bending are disregarded in a first approximation \cite{lond}. 

The work is organized as follows. Section \ref{sec2} gives the details of the theoretical model and
formalism.
In a first stage (Sec.\ \ref{sec3})  the analysis of results is performed assuming
the field can be tuned independently in both arms of the junction. This allows a 
more clear understanding of the physical behavior.
The uniform (homogenous) field is
then considered in Sec.\ \ref{sec4}
for 
varying  
wire widths and Rashba strengths. The robustness of the oscillation suppression 
for finite biases and considering lateral superconductors of different 
pairing-gap phases is discussed in
Sec.\  \ref{sec5}.
Finally,  Sec.\ \ref{sec6} summarizes the conclusions of the work.

\section{Theoretical model and formalism}
\label{sec2}

We consider nanowires containing all the ingredients to create topological phases and magnetoconductance oscillations. Hybrid nanowires that combine s-wave superconductivity, Rashba interaction, and an external magnetic field in a parallel orientation are known to sustain a topological phase if a critical magnetic field is exceeded \cite{Oreg}. s-wave hybrid superconductivity is achieved by proximity coupling a semiconductor nanowire with a superconductor while  Rashba interaction is a property of the semiconductor nanostructure due to the 
confinement asymmetry of the nanowire. 
On the other hand, for magnetoconductance oscillations we need a two-dimensional NS junction with orbital effects in the normal side but not in the superconductor side. This condition is automatically met with true metallic superconductors that do not allow the penetration of magnetic fields to its interior. However, with hybrid nanowires superconductivity is obtained by proximity and a more clever arrangement is needed in order to avoid orbital effects in the gapped region of the hybrid superconducting section, such as in Fig.\ \ref{F1}.

\subsection{The model}

A 2D junction, with $x$ and $y$ the longitudinal (parallel to transport) and transverse coordinates, respectively,   
is described by the following Bogoliubov-de Gennes Hamiltonian 
\begin{equation}
\mathcal{H}_{BdG}=\mathcal{H}_{W}+\mathcal{H}_{Z}+\mathcal{H}_{R}+\mathcal{H}_{0}\;,
\end{equation}
where  $\mathcal{H}_{W}$ contains the kinetic and confinement potential terms
\begin{equation}
\mathcal{H}_{W}=\left( \frac{p_x^2+p_y^2}{2m}+ V(y) - \mu \right)\tau_z   \; ,
\end{equation}
$\mathcal{H}_{Z}$ is the Zeeman term,
\begin{equation}
\mathcal{H}_{Z}=\Delta_B(x)\, \vec{n}(x)\cdot\vec{\sigma}\; ,
\end{equation}
$\mathcal{H}_{R}$ is the Rashba spin-orbit term,
\begin{equation}
\mathcal{H}_{R}=\frac{\alpha(x)}{\hbar}\left(p_x \sigma_y -  p_y \sigma_x \right)\tau_z \; ,
\end{equation}
and, finally, $\mathcal{H}_0$ is the superconductor pairing term
\begin{equation}
\mathcal{H}_0=\Delta_0(x)\, \tau_x\; .
\end{equation}
The potential $V(y)$ models the transversal confinement, with zero value inside the nanowire and infinite outside. 
In the Zeeman term, $\vec{n}$ gives the orientation of the field,
along $z$ and $x$  
for the N and S sides, respectively. The intensities of Zeeman, Rashba and 
pairing interactions are given by $\{\Delta_B(x), \alpha(x), \Delta_0(x)\}$ and 
they may take different constant values in the N and S sides, i.e.,
$\{\Delta_B^{(N)}, \alpha^{(N)}, \Delta_0^{(N)}\}$ and
$\{\Delta_B^{(S)}, \alpha^{(S)}, \Delta_0^{(S)}\}$.
Orbital magnetic effects are included  for perpendicular fields through the substitution $p_x\rightarrow p_x-\hbar y /l_z^2$ as, e.g., in Ref.\ \cite{Os15}, with the magnetic length defined by the 
perpendicular component of the field $B_z$ as 
$l_z^2=\hbar c/eB_z$. The Zeeman parameter $\Delta_B$ is related to the modulus of the magnetic 
field $B$ as $\Delta_B=g^* \mu_B B/2$, with $g^*$ the effective gyromagnetic factor
and $\mu_B$ the Bohr magneton.

We shall obviously have $\Delta_0^{(N)}=0$, i.e., no pairing gap in the normal side, and the model also assumes a priori that the two Zeeman intensities $\Delta_B^{(N)}$ and $\Delta_B^{(S)}$ can be varied independently. The latter is only for discussion purposes since a more realistic situation with 
a uniform field requires $\Delta_B^{(N)}=\Delta_B^{(S)}$ (Sec.\ \ref{sec4}). Also, different Rashba intensities ($\alpha^{(N)}$, $\alpha^{(S)}$) represent a modification of the vertical asymmetry,
e.g., with a gate, of one of the junction sides with respect to the other.   

\subsection{Resolution method} 

For a given energy $E$ we study the solutions of the equation
\begin{equation}
\left(
\mathcal{H}_{BdG}-E
\right)\, \Psi(x,y,\eta_\sigma,\eta_\tau) = 0\; ,
\end{equation}
where $\eta_\sigma,\eta_\tau$ are the spin and isospin (electron-hole) 
discrete variables.
The solution is obtained by means of the numerical method presented in \cite{Osca16}. 
The algorithm is based on the quantum-transmitting-boundary method and,
in essence, 
provides a way of matching two different sets of asymptotic solutions
characterized as superpositions  of complex-$k$ plane waves.
Advantages of this approach are the high spatial resolution and the 
inclusion of large numbers of evanescent modes
without requiring large 2D grids, i.e, without
exceedingly large computational costs.

In the leads, at both sides of the junction, the wave function can be expressed as a superposition
of plane waves, where the wavenumber $k$ may be real or complex and is a characteristic of the fully translationally  invariant wire \cite{Serra}.
For each lead, labeled as (contact) $c=N,S$ the wavefunction thus reads
\begin{equation}
\Psi (x,y,\eta_\sigma,\eta_\tau)=\sum_{\alpha,n_\alpha} 
d_{n_\alpha}^{(\alpha,c)}\, 
e^{i k_{n_\alpha}^{(\alpha,c)}x}\, \phi_{n_\alpha}^{(\alpha,c)}(y,\eta_\sigma,\eta_\tau)\;,
\end{equation}
where $\alpha=i,o$ labels the input/output condition of the mode and 
$n_\alpha=1,2,\dots$ 
the mode number. The wavenumbers $k_{n_\alpha}^{(\alpha,c)}$ and the transversal components  $\phi_{n_\alpha}^{(\alpha,c)}$ are obtained 
solving a transformed version of the BdG equation for the contacts, recast as a non-Hermitian
eigenvalue problem for the wavenumbers \cite{Osca16}.
It is assumed that a large set of them is known.
The coefficients $d_{n_\alpha}^{(\alpha,c)}$, the channel amplitudes, 
are obtained from the matching equations. Note that the energy is used here as a parameter that is fixed selecting the proper input channel. Finally, the conductance is calculated from the wavefunction as
\begin{equation}
\frac{dI}{dV}(E)=\frac{e^2}{h}\, \left[\, N(E)-P_{ee}(E)+P_{eh}(E) 
\rule{0cm}{0.3cm}
\,\right]  \;,
\end{equation}
where 
\begin{eqnarray}
P_{ee}(E) &=& \sum_{n_o\,\eta_\sigma}
\left| d_{n_o}^{(o,N)} \right|^2\,
\int{
dy\,  \left| \phi_{n_o}^{(o,N)}(y,\eta_\sigma,\Uparrow) \right|^2
}\; ,\\
P_{eh}(E) &=& \sum_{n_o\,\eta_\sigma}
\left| d_{n_o}^{(o,N)} \right|^2\,
\int{
dy\,  \left| \phi_{n_o}^{(o,N)}(y,\eta_\sigma,\Downarrow) \right|^2
}\; 
\end{eqnarray}
are the normal (electron-electron) and Andreev (electron-hole) reflection
probabilities.
Note that 
$d_{n_o}^{(o,N)}\phi_{n_o}^{(o,N)}(y,\eta_\sigma,\eta_\tau)$ 
is the spinor amplitude in a particular electron ($\eta_\tau=\Uparrow$)
or hole ($\eta_\tau=\Downarrow$) channel with spin ($\eta_\sigma=\uparrow,\downarrow$)
and corresponding to output towards the N lead.

\begin{figure}[t]
\centering\resizebox{0.355\textwidth}{!}{
\includegraphics[width=\columnwidth,clip=true]{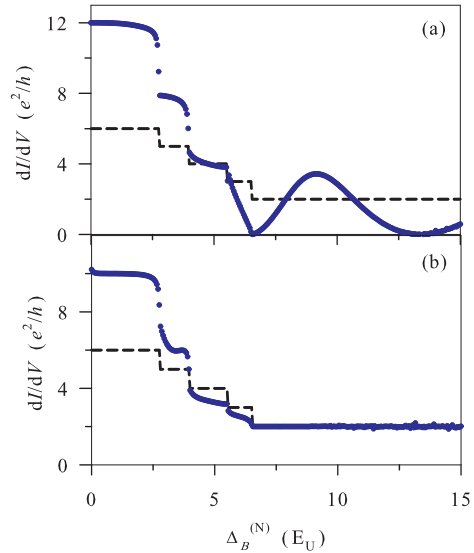}%
}
\caption{Magnetoconductances as a function of $\Delta_B^{(N)}$ for a fixed $\Delta_B^{(S)}$ (blue dots). The number of propagating modes in the N wire 
is given by the dashed line. Panel a) corresponds to a trivial superconductor 
($\Delta_B^{(S)}= 6\, E_U$) and panel b) to a topological superconductor 
($\Delta_B^{(S)}=8\, E_U$).
The remaining parameters are $E=0$, $\alpha^{(S)}=\alpha^{(N)}=2\, E_U L_U$, $\Delta_0^{(S)}=0.7\, E_U$, $\mu=70\, E_U$.
The ratio $\Delta_0^{(S)}/\mu$ is small enough ($\approx 0.01$)
to  ensure the validity of the so-called BTK regime \cite{btk}. The smallest
critical  magnetic field of the hybrid wire, Eq.\ (\ref{eq11}),
is $\Delta_{B,4}^{(c)}\approx 7\, E_U$. 
}
\label{F2}
\end{figure}

\section{Independently tunable fields}
\label{sec3}

The results discussed below are given in the same unit system
of Ref.\ \cite{Os15}, characterized by 
a length unit $L_U$ and a corresponding energy unit $E_U=\hbar^2/mL_U^2$.
A natural choice is $L_U=L_y$, the transverse width of the 2D stripe
although  below we will also use in some specific cases 
different values for $L_U$ and $L_y$.
With $L_U=150$ nm and $m=0.033m_e$, typical with InAs, it is
$E_U=  0.10$ meV. From $\Delta_B = g^* \mu_B B/2$, with 
$\mu_B$ the Bohr magneton and $g^*=15$ (gyromagnetic factor),
the magnetic field modulus corresponding to a given $\Delta_B$ is
$B=0.23\, (\Delta_B/E_U)\, {\rm T}$.

\subsection{Conductance in trivial and topological phases}

Firstly, we study the magnetoconductance of the junction model 
of Sec.\ \ref{sec2} assuming that the magnetic field in 
the two junction sides (the two arms of the cleaved-edge nanowire) can be tuned independently.
Although this is not very realistic, it is a good starting point from a theoretical point of view as it allows controlling the topological and trivial phases of the 
(hybrid) superconductor without changing the parameters for the normal side. This way, 
any difference between the two situations can be ascribed to the superconductor
phase condition.

Figures \ref{F2}a and \ref{F2}b show the magnetoconductance as a 
function of the magnetic field on the normal side
of the junction $\Delta_B^{(N)}$, while the superconductor side of the 
junctions remains under a constant field $\Delta_B^{(S)}$. The 
figure also shows the evolution of the number of propagating modes in the normal side
(dashed line). Note that, this number decreases with the magnetic field in a stepwise manner, with 
steps of one unit. This behavior is due to our normal contact 
including a non zero Zeeman term (cf.\ Ref.\ \cite{Taka98}). 
We consider a nanowire made of an homogeneous material, hence with a constant Rashba term 
throughout. Together, Rashba and Zeeman terms
split and mix the spin degrees of freedom in a way that full pure Andreev reflection is not achieved for low magnetic fields, but there is some normal reflection too.
As a consequence, the conductance in this limit is less than two times the number 
of active modes.
Full Andreev reflection at vanishing fields is recovered 
when disregarding the Zeeman and Rashba terms (but maintaining the orbital effects). 

\begin{figure}[t]
\centering\resizebox{0.5\textwidth}{!}{
\includegraphics[width=\columnwidth,clip=true]{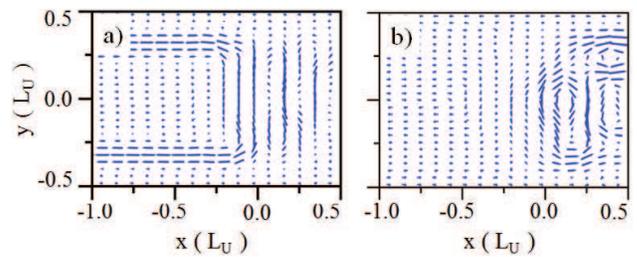}%
}
\caption{Pattern of probability currents for the trivial (a) and topological (b) phases of the superconductor lead.
The parameters (for each panel) are the same as in Fig.\ \ref{F2}.
The incident and reflected currents in b) are difficult to distinguish because 
of the vortex scale dominating the normalization of the plot.
}
\label{F6}
\end{figure}

In Fig.\ \ref{F2}a the superconductor is kept in a trivial phase with  
$\Delta_B^{(S)}$ below a critical value, while in Fig.\ \ref{F2}b it is in a
topological phase because of the larger $\Delta_B^{(S)}$. 
While both figures are similar for weak magnetic fields, clear conductance oscillations arise for strong fields when the superconductor is in the trivial phase. These oscillations are of the same kind and share the same origin of those discussed in Ref.\ \cite{Taka98}. In a semiclassical image the incident electrons are reflected as holes by the junction but the orbital effect  bends their trajectories towards the
junction again. This process is repeated along the transverse boundary 
creating  a pattern that may enhance or suppress Andreev reflection against normal reflection. This effect can be seen as causing a complicated vortex structure in the junction (Fig.\ \ref{F6}a). In the particular case of Fig.\ \ref{F2}a
the oscillations are seen only in the plateau of two active modes; however, higher chemical potentials $\mu$ allowing more active modes for strong magnetic fields would enable oscillations at different incident conditions.

On the other hand, when the superconductor is in the topological phase (Fig.\ \ref{F2}b)  the oscillations are fully suppressed in the two-mode plateau. 
This occurs because the vortex structure in the junction is dramatically 
modified with respect to the trivial case. 
In the topological phase a single dominant vortex is formed at the junction
(Fig.\ \ref{F6}b) irrespective of the strength of the magnetic field, eliminating this way any transverse structure that may enhance or suppress Andreev reflection. Note, too, that in the topological phase only one channel is open to Andreev reflection while the other one is reflected as normal electron-electron back scattering. 
The reason is that 
the topological number of the superconductor is only zero or one, larger values 
being forbidden in the D symmetry class of the Hamiltonian. As mentioned above, the
complete suppression of magnetoconductance oscillations is only achieved 
in the two-mode plateau. Only one channel can be attached to the topological mode of the superconductor while the rest undergo Andreev 
or normal reflection by means of the usual physics of the trivial superconductor.

\subsection{Chemical potential dependence}

\begin{figure}[t]
\centering\resizebox{0.48\textwidth}{!}{
\includegraphics[width=\columnwidth,clip=true]{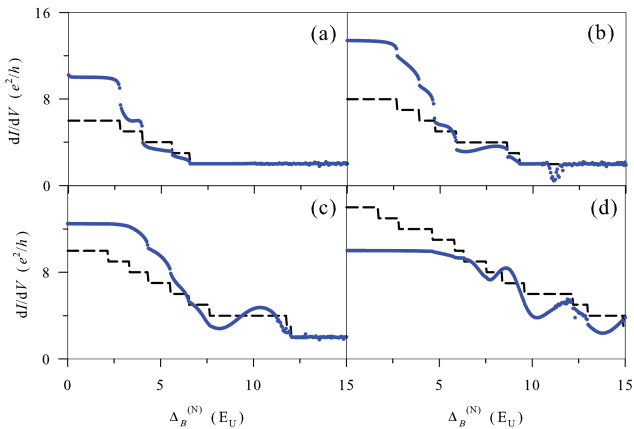}%
}
\caption{Magnetoconductances for different chemical potentials of the normal and superconductor leads $(\mu^{(N)},\mu^{(S)})$.
We have used fixed values $\Delta_B^{(S)}=8\, E_U$ and $\mu^{(S)}=70\, E_U$
(topological superconductor), 
while for each panel $\mu^{(N)}$ is: a) $70.5\, E_U$, b) $105\, E_U$, c) $140\, E_U$ and 
d) $250\, E_U$.
Dashed line and rest of parameters as in Fig.\ \ref{F2}.
}
\label{F3}
\end{figure}

In Fig.\ \ref{F3} we show magnetoconductances of the NS junction when the 
chemical potentials  of the normal and superconductor regions differ. The superconductor region is kept in the topological phase using the same parameters of Fig.\ \ref{F2}b and the changes
in chemical potential 
are only in the normal region. The main effect of increasing the N chemical potential is the displacement to higher magnetic fields of the plateau of two incident modes. Therefore the region of conductance oscillation suppression is consistently displaced to stronger magnetic fields too. Furthermore, for high enough N chemical potentials  
oscillations appear in regions where more than two incident modes are active. This is the usual physics of magnetoconductance oscillations with trivial superconductors  and, as discussed above, the
oscillation suppression affects only the channel that attaches to the topological mode of the superconductor.

\section{Uniform magnetic fields}
\label{sec4}

After having analyzed in the preceding section the topological suppression of
oscillations with two independent tunable fields, we study now the more realistic 
case of the cleaved-edge nanowire in a uniform field (Fig.\ \ref{F1}).
Our model neglects the presence of localized states at the cleavage but it is 
enough to study the magnetoconductance oscillations.
The modulus of the field may be above or below the critical value 
for the S lead. This critical value in presence of a 2D Rashba field 
was discussed in Ref.\ \cite{Os15}, generalizing previously known 
expressions \cite{Oreg,Lutchyn,sanjose}.  
It reads
\begin{equation}
\label{eq11}
\Delta_{B,n}^{(c)}=\sqrt{\left(\mu-\epsilon_n+\frac{m \alpha^2}{2 \hbar^2}\right)^2+\Delta_o^2}
\end{equation}
where 
$\epsilon_n=\hbar^2 \pi^2 n^2/ 2 m L_y^2$ with $n=1,2,\dots$ are the 1D square well 
eigenenergies.

\subsection{Width dependence}

\begin{figure}[t]
\centering\resizebox{0.355\textwidth}{!}{
\includegraphics[width=\columnwidth,clip=true]{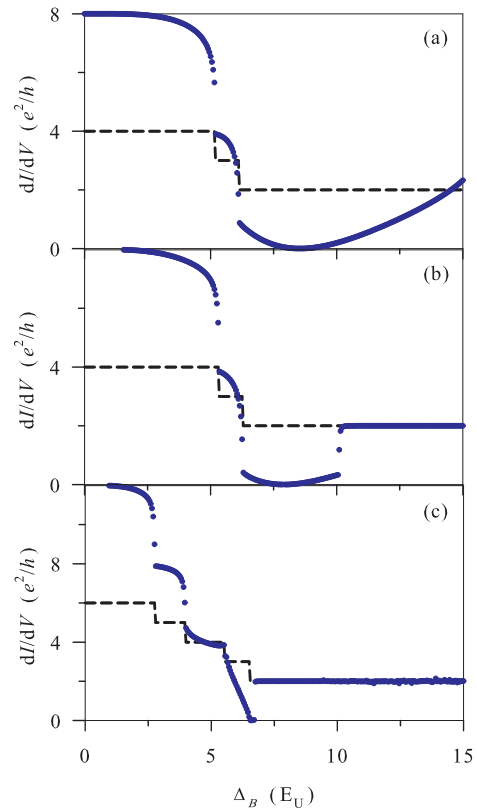}%
}
\caption{Magnetoconductances of the cleaved-edge wire in uniform field
(Fig.\ \ref{F1}), using $\Delta_B=\Delta_B^{(S)}=\Delta_B^{(N)}$,
for selected wire widths $L_y$: a) $0.7\, L_U$, b) $0.73\, L_U$, c) $1.0\, L_U$.
Dashed line and rest of parameters as in Fig.\ \ref{F2}.
}
\label{F4a}
\end{figure}

From Eq.\ (\ref{eq11}) it is clear that 
we can adjust the critical field by changing the width of the nanowire. In this manner we can choose in which phase of the hybrid superconductor we find the plateau of two incident modes, where we expect to recover the behavior discussed in the preceding section. 
For instance, in Fig.\ \ref{F4a}a the nanowire is narrow enough  to find the superconductor only in its trivial phase, while in Fig.\ \ref{F4a}c the nanowire is wide enough to find the plateau of two incident modes already in the topological region. We can also find an intermediate width in which oscillations are present at the beginning of the plateau and perfect suppression is seen in the rest of the plateau (Fig.\ \ref{F4a}b). Both regimes are separated in an abrupt way by the critical magnetic field. This is yet another proof of the topological suppression of 
magnetoconductance oscillations as it appears exactly at the expected critical value.

\subsection{Dependence on Rashba coupling}

\begin{figure}[t]
\centering\resizebox{0.35\textwidth}{!}{
\includegraphics[width=\columnwidth,clip=true]{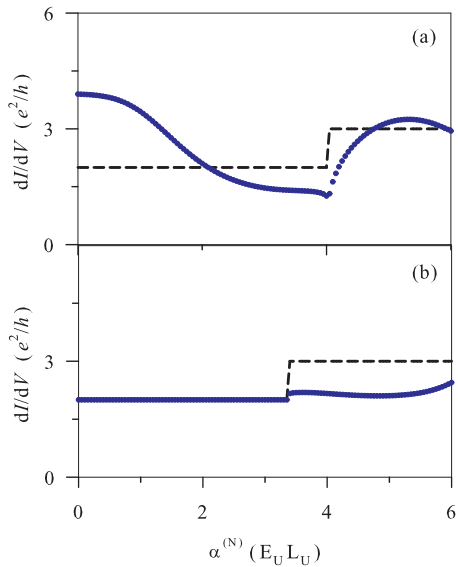}%
}
\caption{Magnetoconductance as a function of the Rashba coupling in the normal side $\alpha^{(N)}$.
We have used fixed values $\Delta_B=\Delta_B^{(N)}=\Delta_B^{(S)}=8\, E_U$, $\alpha^{(S)}=2 E_U L_U$ and $\Delta_0^{(S)}=0.7 E_U$.
The superconductor is always in the trivial phase for panel a) where $\mu^{(S)}=60\, E_U$
while it is in the topological phase for panel b) with $\mu^{(S)}=70\, E_U$. The rest of parameters are the same of Fig.\ \ref{F2}.
}
\label{F4b}
\end{figure}

Another option to avoid using spatial variations of the magnetic field strength is 
tuning the intensity of the Rashba coupling.
It is well known from spintronics that $\alpha$ can be tuned 
using external electric gates \cite{Nit97}. 
If the superconductor is in a trivial phase we observe variations in the conductance when changing the Rashba coupling in the normal side of the junction (Fig.\ \ref{F4b}a). 
On the contrary, if the superconductor is in the topological phase the conductance remains stuck at the quantized value provided that only two incident modes are active, as shown in Fig.\ \ref{F4b}b. We are assuming the existence of a gate that allow us 
to tune the Rashba coupling in the normal side of the junction independently of the superconductor side. With large enough $\alpha$'s  additional incident modes are active, thus generating the magnetoconductance variations observed
at the right end of Fig.\ \ref{F4b}b.

Besides the two-mode restriction, it is important to also take into 
account that the value of the critical field, Eq.\ (\ref{eq11}), 
changes with  $\alpha$ and the chemical potential in the superconductor side. In particular in Fig.\ \ref{F4b} we have changed 
the overall chemical potential in the nanowire to control its phase while maintaining the homogeneous external magnetic field fixed. 

\section{Robustness and generality}
\label{sec5}

A relevant question is how robust is the suppression effect discussed in this work. Suppression of magnetoconductance oscillations occurs around zero energy,
corresponding to vanishing potential bias on the NS junction.
Moving away from zero energy (non zero bias) will eventually give rise to oscillations again. The suppression, however, 
is not a single-energy phenomenon fading away  
with infinitesimal energies but survives for energies within a finite range.
The region around zero in which the suppression occurs gets wider with an increasing superconducting gap. On the other hand, it is known that 
trivial superconductors stop providing Andreev reflection for large gaps, and therefore any related magnetoconductance oscillations are also quenched for large gaps. 
Nevertheless, Andreev reflection with oscillation suppression is more robust 
and easier to find with large superconductor gaps in the topological phase of the superconductor.

We believe magnetoconductance suppression is a general property of any kind of topological superconductor. We have actually checked that with a p-wave superconductor in the topological phase \cite{Wim10} the same kind  of suppression of magnetoconductance 
variations is found. However, the main difference between p-wave and s-wave
cases is that 
with p-wave superconductors it is no longer appropriate to think of 
topological oscillation
suppression since oscillations are never present \cite{Yan14}.
This is because no Andreev reflection is found in the trivial phase of the p-wave. Andreev reflection is only found in the topological phase 
and therefore Takagaki-like oscillations are never found with p-wave superconductivity. 

\begin{figure}[t]
\centering\resizebox{0.5\textwidth}{!}{
\includegraphics[width=\columnwidth,clip=true]{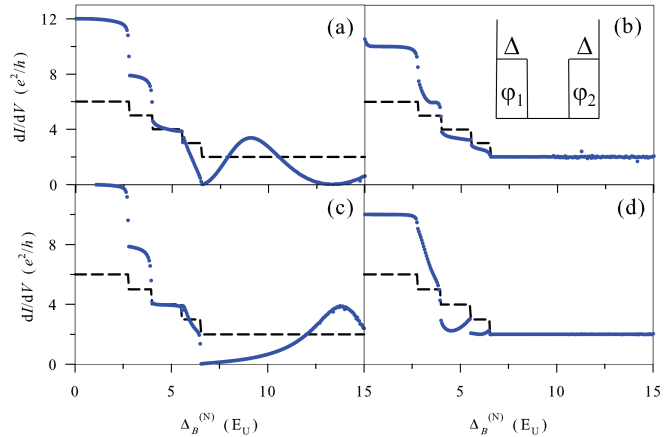}%
}
\caption{Magnetoconductance when the hybrid nanowire is proximity coupled 
to two different superconductors with phases $\phi_1$ and $\phi_2$ (phase
difference $\phi\equiv\phi_1-\phi_2$) sketched in the inset
to panel b. Each superconducting region has a width
$0.2\, L_U$, while the width of the intermediate region
is $0.6\, L_U$.
The 
different panels are for 
a) $\Delta_B^{(S)}=6\, E_U$ (trivial phase), $\phi=0$, 
b) $\Delta_B^{(S)}=8\, E_U$ (topological phase), $\phi=0$,  
c) $\Delta_B^{(S)}=6\, E_U$ (trivial phase), $\phi=\pi$ and 
d) $\Delta_B^{(S)}=8\, E_U$ (topological phase), $\phi=\pi$.
Dashed line and rest of parameters are as in Fig.\ \ref{F2}.
}
\label{F5}
\end{figure}

\subsection{Lateral S phases}

We next ask ourselves what happens to the magnetoconductance oscillations 
if the hybrid nanowire is laterally coupled to two different superconductors with different phases. One superconductor will be proximity coupled at one side of 
the wire while the other is in contact with the other side in the transverse direction. The middle region of the hybrid nanowire will remain 
normal. 
The resulting transverse structure
is sketched in Fig.\ \ref{F5}b. 
As shown in Fig.\ \ref{F5}a, it is not needed that the superconductor pairing  extends throughout the nanowire to obtain magnetoconductance oscillations, provided the superconductors are in the trivial phase. Furthermore, if the superconductors are in the topological phase we recover the same kind of oscillation suppression already seen for the full superconducting nanowire.  Remarkably, in Fig.\ \ref{F5}c we can see how a phase difference between both superconductors changes the shape of the conductance oscillations in the trivial phase, displacing the positions of maxima and minima. Despite these changes in the trivial phase, in panel \ref{F5}d we can check again the robustness of the oscillation suppression, 
that is now resilient to superconductor phase changes in the two-mode plateau.

\section{Conclusions}

\label{sec6}

In this work we have shown how a superconductor in topological phase suppresses the magnetoconductance oscillations of NS junctions when 
the N wire sustains only two active modes.
These oscillations are present 
under the same conditions but for the superconductor in the trivial phase. This result is due 
to the formation of a large vortex of probability current 
at the junction when the S wire enters the topological phase.
We have analyzed the phenomenon from a theoretical point of view, assuming independently tunable fields on both junction sides, and we also
studied more realistic scenarios of cleaved-edge wires in uniform fields. 
In the latter case it is possible to tune the suppression of the oscillations
by changing the width of the nanowire.

We have proved the robustness of the oscillation suppression provided the superconductor topological phase is achieved. Similar oscillations can be found (or suppressed) tuning 
the Rashba coupling with an external electric field. With p-wave superconductors the same physics is found
in the topological phase, although magnetoconductance oscillations are absent in the trivial phase and one can not properly speak of oscillation suppression. 
Finally, we have seen how the oscillations depend on the phase difference between two trivial superconductors proximity coupled with the nanowire in the transverse direction. In this case too, when the superconductors are in topological phase
the suppression is robust and independent of their relative phase difference.

\begin{acknowledgments}
This work was funded by MINECO-Spain (grant FIS2014-52564),
CAIB-Spain (Conselleria d'Educaci\'o, Cultura i Universitats) and 
FEDER.
\end{acknowledgments} 

\bibliographystyle{apsrev4-1}
\bibliography{ArticleA6LSbib}

\end{document}